\documentclass[fleqn,10pt]{wlscirep}
\usepackage[utf8]{inputenc}
\usepackage[T1]{fontenc}

\usepackage{placeins}
\usepackage{xparse}
\DeclareMathAlphabet{\mathcal}{OMS}{cmsy}{m}{n}

\renewcommand{\v}[1]{\mathbf{#1}}
\newcommand{\pd}[2]{\frac{\partial #1}{\partial #2}}
\newcommand{\s}[1]{\mathrm{#1}}
\renewcommand{\mod}[1]{|#1|}
\NewDocumentCommand{\evalat}{sO{\big}mm}{%
	\IfBooleanTF{#1}
	{\mleft. #3 \mright|_{#4}}
	{#3#2|_{#4}}%
}

\title{Anderson localisation in steady states of microcavity polaritons}

\author[1]{Thomas J. Sturges\thanks{sturges.tom@gmail.com}}
\author[2]{Mitchell D. Anderson}
\author[1]{Adam Buraczewski}
\author[2]{Morteza Navadeh-Toupchi}
\author[2]{Albert F. Adiyatullin\thanks{Current affiliation: Department of Physics, MIT-Harvard Center for Ultracold Atoms, and Research Laboratory of Electronics, Massachusetts Institute of Technology, Cambridge, Massachusetts 02139, USA}}
\author[2]{Fauzia Jabeen}
\author[2]{Daniel Y. Oberli}
\author[2]{Marcia T. Portella-Oberli}
\author[1]{Magdalena Stobi{\'n}ska}

\affil[1]{Institute of Theoretical Physics, University of Warsaw, ul. Pasteura 5, 02-093, Warsaw, Poland}
\affil[2]{Institute of Physics, School of Basic Sciences, Ecole Polytechnique F{\'e}d{\'e}rale de Lausanne, 1015 Lausanne, Switzerland}

\begin{abstract}
We present an experimental signature of the Anderson localisation of microcavity polaritons, and provide a systematic study of the dependence on disorder strength. We reveal a controllable degree of localisation, as characterised by the inverse-participation ratio, by tuning the positional disorder of arrays of interacting mesas. This constitutes the realisation of disorder-induced localisation in a driven-dissipative system. In addition to being an ideal candidate for investigating localisation in this regime, microcavity polaritons hold promise for low-power, ultra-small devices and their localisation could be used as a resource in quantum memory and quantum information processing.

\end{abstract}
\begin{document}

\flushbottom
\maketitle

\thispagestyle{empty}

\section*{Introduction}

The localisation of diffusive waves due to an underlying disorder is pervasive throughout nature. A Nobel-prize winning description of the localisation of single-particle electronic wavefunctions in periodic lattices was provided by Anderson in 1958 \cite{Anderson1958}. The essence being that for sufficiently strong on-site disorder, the eigenstates transition from extended Bloch states to exponentially localised. Since then, localisation has been realised in several distinct physical systems such as photonic waveguides \cite{Segev2013}, microwaves \cite{Dalichaouch1991, Chabanov2000} and sound waves \cite{Weaver1990}. True Anderson localisation of matter waves remained elusive for a long time, and was only realised in an atomic Bose-Einstein condensate in 2008 \cite{Billy2008}. In addition, the question of whether single particle states remain localised in the presence of inter-particle interactions was eventually resolved positively \cite{Basko2006, Gornyi2005}, and has led to the field of many-body localisation \cite{Abanin2017}.

In general, localisation prevents the system from reaching the equilibrium state of the corresponding non-disordered system. The most striking example can be found in closed systems of cold atoms, where a random disorder potential prevents the system from reaching thermal equilibrium \cite{Billy2008, Schreiber2015, Choi2016}. In contrast to closed regimes, we are interested in localisation in the steady states of driven-dissipative systems, the solutions that arise from the balance of flows from driving and dissipation. It is not immediately obvious if the disorder-induced Anderson localisation has a role to play in this regime, and so it is an interesting question to explore experimentally.

Microcavity polaritons are an ideal test-bed for this. They are hybrid bosonic quasiparticles that arise due to the strong interaction between microcavity photons and quantum well excitons \cite{Kavokin2017}. They have lifetimes on the order of picoseconds, and thus the polariton population must be continually replenished by a laser source in order to form a steady state. Due to advanced fabrication techniques, the potential landscape experienced by the polaritons can be exactly engineered, and their distribution can be directly mapped by measuring the photoluminescence. We note that microcavity polariton localisation has been reported in a few different contexts such as: nonlinear-induced localisation when resonantly driving one micropillar of an interacting dimer \cite{Rodriguez2018, Abbarchi2013}; gain-induced localisation in finite width excitation spots \cite{Roumpos2010}, a consequence of a finite polariton lifetime; and localisation in flat bands where the polaritons have an infinite effective mass \cite{Baboux2016}.  However, to date, no systematic study of the effect of disorder-induced Anderson localisation has been performed.

Here we experimentally demonstrate a signature of the Anderson localisation of exciton-polaritons in two-dimensions. To do this, we study the steady-state polariton distribution under non-resonant excitation in a set of eight hexagonal lattices with increasing levels of static disorder. The static off-diagonal disorder is introduced by adding a random displacement to each lattice site with a controllable maximum amplitude. The localisation is characterised by the inverse participation-ratio, and is shown to monotonically increase as a function of disorder strength. The experimental results are supported by numerical simulations of a Gross-Pitaevskii equation, with which we also explore the effect of the polariton nonlinearity.

\section*{Results}

\begin{figure*}[t!]
	\centering
	\includegraphics[width=0.8\textwidth]{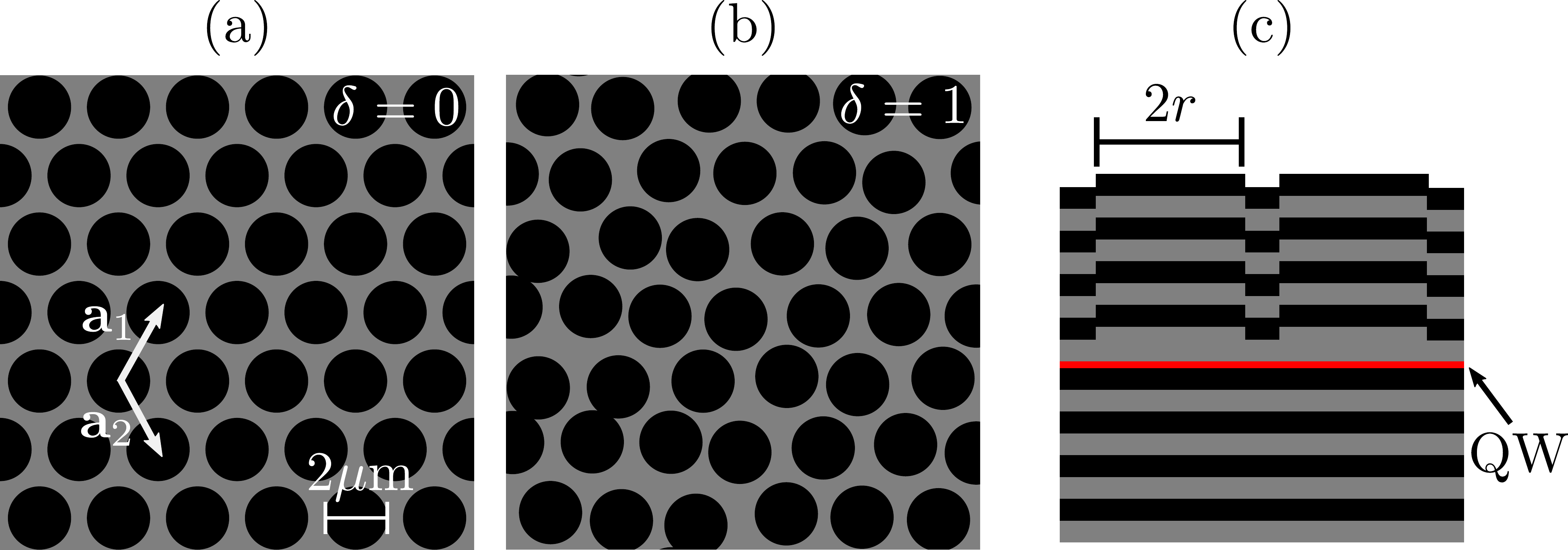}
	\caption{\textbf{Schematic of the system geometry}. A bird's eye view of a region of the device is shown for (a) no disorder $\delta=0$ and (b) maximum disorder $\delta=1$, along with the lattice vectors $\v{a}_{1,2} = (a/2)(1,\pm\sqrt{3})$. (c) A side-view of the active region with the quantum well (QW) shown in red.} 
	\label{fig:schematic}
\end{figure*}

\begin{figure*}[t]
	\centering
	\includegraphics[width=\textwidth]{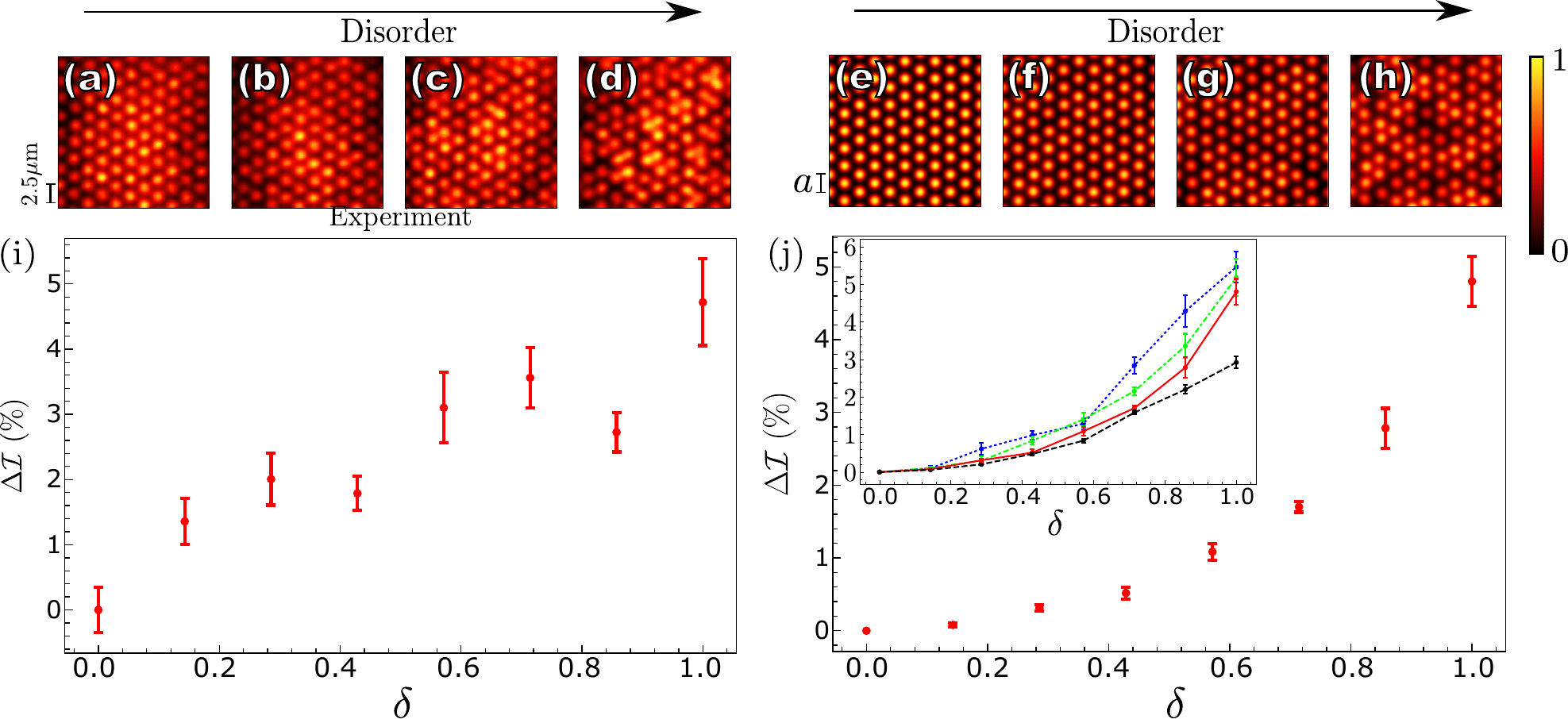}
	\caption[]{\textbf{Signature of localisation.} Figures (a)-(d) show the real-space photoluminescence images under weak (P = 100 mW) nonresonant excitation, whereas figures (e)-(h) show the polariton density in the numerical model. In both cases the results are normalized to account for the Gaussian pump distribution (see Methods) and the disorder levels are (a,e) $\delta=0$, (b,f) $\delta=0.284$, (c,g) $\delta=0.572$, (d,h) $\delta=1$. In the absence of disorder we see a more homogeneous distribution among the mesas, whereas disorder induces the onset of patches of localisation. Also shown are plots that reveal how the IPR increases with disorder for (i) experiment, and (j) theory. The polariton nonlinearity in (j) is $g=g_0= 2.4\times 10^{-3} \s{meV}.\mu\s{m}^2$. The inset of (j) shows the result for $g=-g_0$ (blue dotted line), $g=0$ (green dot-dashed line), $g=g_0$ (red solid line), and $g=10g_0$ (black dashed line). The lines are just a guide for the eye. In the experimental figures the cavity-exciton detuning is $2$meV. The simulation parameters are $m = 5\times10^{-5} m_e$, $\hbar R =  0.4 \s{meV}.\mu\s{m}^2$, $\hbar \gamma  =  0.5 \s{meV}$, $\hbar g  = 2.4\times 10^{-3} \s{meV}.\mu\s{m}^2$, $V_0 = 9 \s{meV}$, $\hbar \gamma_R  = 2 \s{meV}$, and $P_0 = 2 \gamma_R \gamma / R$. Here, $V_0$ is the maxima of the trapping potential, which we model as a radially symmetric sigmoid function for each mesa.}
	\label{fig:main}
\end{figure*}

Trapping potentials for the polaritons are designed by patterning the microcavity with mesas \cite{Daif2006, Jacqmin2014, Milicevic2015}, circular elongations of the cavity spacer that locally alter the cavity detuning. See figure \ref{fig:schematic} for a schematic of the device. We arrange the mesas in a hexagonal lattice, and close enough that the quantised modes of the individual mesas \cite{Daif2006} hybridise into extended modes \cite{Adiyatullin2017} whose features depend on the lattice geometry. The polariton population is fed by a reservoir of excitons created by the non-resonant excitation of a continuous-wave laser. From the photoluminescence images of the perfect hexagonal arrays, Figure \ref{fig:main}(a), we see the confinement of polaritons predominantly within the mesas and an approximately homogeneous distribution among them. We note that the coupling between the mesas was evidenced in the spectral distribution of the polaritons (see Supplementary Fig. S3) where one can see the periodic Bloch bands of the hexagonal lattice.

To introduce an off-diagonal disorder we shift the cartesian coordinates of each mesa by a random displacement in the range $d[-\delta,\delta]$, where $0 \leq \delta \leq 1$ parameterises the amount of disorder, and $d=0.25\mu\s{m}$ is the maximum possible displacement for the maximum disorder $\delta=1$. This positional disorder
modifies the eigenstates of the system from Bloch states
towards spatially separated patches of localisation. The effect of the disorder on the localisation, or clustering, of the polariton population can be seen by eye in the photoluminescence images, Figures \ref{fig:main}(a-d). To obtain a quantitative measure of the amount of localisation we calculate the inverse-participation ratio (IPR), which is functionally similar to imbalance measures used in cold atom experiments \cite{Schreiber2015, Choi2016}, although it is not a site specific measure. The IPR has previously been used to quantify Anderson localisation in photonic systems \cite{Schwartz2007} and is also applicable here. In essence it is a measure of inhomogeneity, and for a homogeneous distribution it is equal to unity. Thus as the onset of Anderson localisation causes some mesas to contribute more significantly to the total photoluminescence, the IPR increases also. To exploit this measure, we first normalise the data to account for the Gaussian background (see Methods), and then calculate the average occupation $I_n = \int_{\s{mesa}_n} \mod{\psi}^2 \s{d}\v{r} $ of each mesa (labelled by $n$). We then obtain the IPR as

\begin{equation}\label{IPR}
\mathcal{I} =  N \left( \sum_{n=0}^N I_n^2 \right)  / \left( \sum_{n=0}^N I_n \right)^2.
\end{equation}

Figure \ref{fig:main}(i) shows the percentage change in the IPR from no disorder ($\delta=0$). We clearly see the increase in IPR with increasing disorder $\delta$, which signals the onset of localisation. The error bars correspond to the standard error after repeating the experiment on 12 different regions sampled from a larger lattice for each disorder strength. In addition, similar results have been reproduced for several different laser powers (see Supplementary Figure S1).

We successfully modelled the experimental results with a generalised Gross-Pitaevskii equation \cite{Wouters2007} describing the evolution of the polariton wavefunction $\psi(\v{r},t)$,

\begin{equation}\label{GPE}
i \hbar \pd{\psi}{t} = \left[ -\frac{\hbar^2 \nabla^2}{2 m} + \frac{i \hbar}{2}\left( R n_R - \gamma \right) + \hbar g \mod{\psi}^2 + V\right] \psi,
\end{equation} 

where $m$ and $\gamma$ are the effective mass and decay rate of the polaritons, $g$ is the strength of polariton-polariton interactions, $R$ is the reservoir-polariton exchange rate, and $V(\v{r})$ is the potential landscape defined by the lattice. The reservoir $n_R(\v{r},t)$ is described by the rate equation

\begin{equation}
\pd{n_R}{t} = -\left( \gamma_R + R \mod{\psi}^2 \right) n_R + P,
\end{equation}

where $\gamma_R$ is the decay rate of the reservoir. The reservoir is populated by the continuous-wave pump $P(\v{r})$ which we model as a Gaussian with amplitude $P_0$.  We use the Runga-Kutta method of fourth order to evolve the dynamics until a stationary solution is achieved (approximately $50\s{ps}$). In figures \ref{fig:main}(e-h) we show the results of these simulations for a periodic and increasingly disordered arrays. We can see the onset of patches of localisation when a disorder is introduced. We calculate the IPR using equation \ref{IPR} in much the same way as is done for the experimental data, and plot the results in Figure \ref{fig:main}(j).

In the inset of Figure \ref{fig:main}(j) we show calculations performed for different polariton-polariton nonlinearities. The linear regime is shown by the green dot-dashed line. In the present experiment we are working in a weak nonlinearity regime, which we model with a small $g$ in the simulations (red solid line). Nonetheless we observe that the positive interaction acts to suppress the localisation; see also the black-dashed line where we increase the non-linearity ten-fold. In addition we show that a negative interaction acts to enhance the localisation (blue dotted line). Although a negative $g$ is not possible with our experimental setup, we include this simulations result as a point of interest for the reader. Such a regime could be accessed with spinor condensates tuned near the Feshbach resonance \cite{Takemura2014}.

\begin{figure*}[t!]
	\centering
	\includegraphics[width=\textwidth]{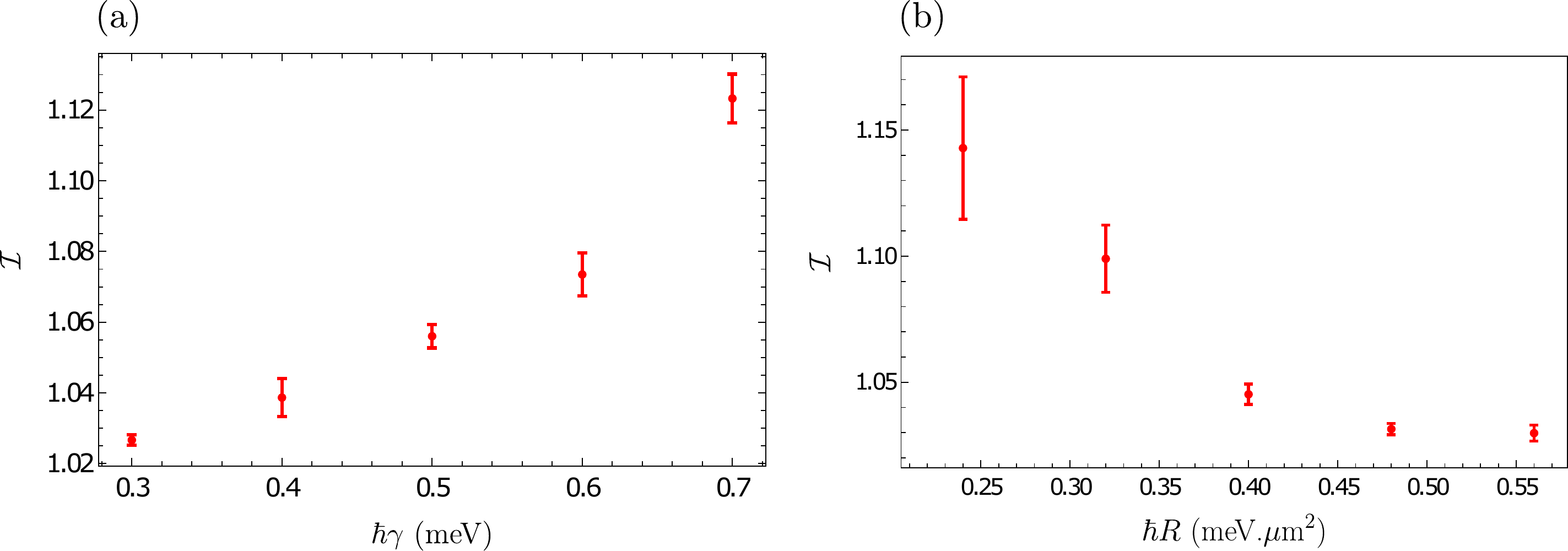}
	\caption{\textbf{Tuning the localisation through system parameters}. Dependence of the IPR on (a) the polariton lifetime, and (b) the polariton-reservoir exchange rate. In both figures $\delta=1$. The other simulation parameters are given in Figure \ref{fig:main}.} 
	\label{fig:tuning_param}
\end{figure*}

In Figure \ref{fig:tuning_param} we provide some further theoretical analysis, demonstrating the ability to tune the localisation through additional system parameters.  For example, simply by varying the detuning of the cavity via the spacer width we can vary e.g.\ the polariton lifetime. As we see in figure \ref{fig:tuning_param}(a), decreasing the lifetime of the polaritons results in an enhanced localisation. As another example we can modify the reservoir-polariton coupling, which could be accomplished e.g. by enhancing phonon mediated relaxation of excitons into polariton states. In this way we can suppress the localisation by effectively increasing the driving of the polaritons, as shown in figure \ref{fig:tuning_param}(b).

\section*{Discussion}

Our results demonstrate a signature of Anderson localisation in the steady states of microcavity polaritons. In general we observe a convincing qualitative agreement between the experimental and numerical results. In both cases the IPR increases monotonically with disorder, and by a similar magnitude, which signals the onset of localisation. The difference between the rates of increase of IPR is attributed to intrinsic disorder of the experimental samples, as well as slightly different data processing algorithms. In this work we were operating in a regime of small nonlinearity where interactions acted to weakly suppress the localisation.

Our results offer a signature of disorder-induced localisation in the steady states of driven-dissipative systems, a regime quite  different to the prototypical case of closed systems. We believe that the phenomenology reported here should be generally observable in other driven-dissipative systems and may open a new chapter for basic science explorations. In addition, since both polaritons and localisation are foreseen to have a high potential for applications in optoelectronic devices and quantum information respectively, such a robust and controllable phenomenon could be of use in novel devices.

To advance our work, it would be interesting to investigate localisation of strongly interacting polaritons. For example, one could explore the interplay between the disorder-induced localisation seen herein and effects such as nonlinear localisation observed for strongly driven microcavities \cite{Rodriguez2018}. Moreover with larger polariton densities it may be possible to extract signatures of many-body localisation, which in a crude sense is the persistence of Anderson localisation in the presence of many-body interactions. One could also explicitly examine the role of localisation in driven-dissipative systems for preserving memory of initial conditions. For example, by preparing initially imbalanced population distributions, with a highly inhomogeneous pumping, and then switching on a homogeneous pumping to see if a signature of the initial state perseveres.

\section*{Methods}

\subsection*{Experimental Methods}

We study a GaAs $\lambda$-microcavity made of $24/20$ pairs of GaAs/AlAs, with embedded 8nm, $\s{In}_{0.04}\s{Ga}_{0.96}\s{As}$ quantum well (with an exciton energy of 1.482 eV) which gives a Rabi splitting of 3.1meV. 
We fabricate circular mesas with a radius of $r = 1\mu\s{m}$ by a $6\s{nm}$ local elongation of the cavity spacer, which provides a trapping potential of $9\s{meV}$ for the polaritons, leading to confined quantised modes in the individual mesas \cite{Daif2006}. We arrange the mesas into a hexagonal array with lattice constant $a=2.5\mu\s{m}$ which is sufficient for wavefunction overlap between neighbouring mesas, giving rise to new hybridised modes \cite{Adiyatullin2017}.
To introduce an off-diagonal disorder to the system the $x$ and $y$ coordinates of the mesas are offset by a random value in the range $d[-\delta,\delta]$ where $d=0.25\mu\s{m}$. We consider eight different disorder levels from $\delta=0$ to $\delta=1$ in evenly spaced steps.

We excite the system non-resonantly with a 660nm continuous wave laser focused to a spot size of $25\mu\s{m}$. We measure the photoluminescence of the sample with a collection lens of numerical aperture 0.42NA, and image the real-space integrated energy emission in a CCD.

We are interested in the onset of localisation phenomena as evidenced by the increase in the IPR. Thus we renormalise the experimentally obtained signal by the pump beam profile to eliminate its effect on the IPR. Thus we fit a Gaussian to the image after filtering out higher frequencies, which leaves us with the overall shape of the pump profile, after relaxing from the higher nonresonant energy. We then crop the image to a region within the central pumping region where the signal-to-noise ratio is sufficient. Then we use a peak-finding algorithm to determine the location of the mesas. We then calculate the average intensity $I_n$ of each mesa (labelled by the integer $n$) in a region around these peaks of the same width as the mesas. The IPR can then be calculated from equation (\ref{IPR}).

\subsection*{Numerical methods}

We use the Runge-Kutta method of fourth order to solve equation (\ref{GPE}). We denote the mesa positions as $\v{R}_n = n_1 \v{a}_1 + n_2 \v{a}_2 + d \boldsymbol{\mathcal{S}}_\delta$, where $\v{a}_1 = a(0,1)$ and $\v{a}_2=(a/2)(1,\sqrt{3})$ are the lattice vectors, $a$ is the lattice constant, and $n$ labels the integers $n_1$ and $n_2$. Here, $\boldsymbol{\mathcal{S}}_\delta$ is a random vector whose coordinates are uniform random variables sampled in the range $[-\delta,\delta]$, $0 \leq \delta \leq 1$ parametrises the disorder, and $d=0.25\mu\s{m}$ is the maximum possible displacement in each direction. We take the zero-energy to be $E_0$, the bottom of the polariton band. So the wavefunction $\psi$ is technically the slowly oscillating envelope of the real wavefunction $\Psi = \psi \exp(-i E_0 t/\hbar)$. This is done to minimise numerical errors that can accumulate with fast oscillations. 

We seed the process with a weak initial state $\psi(t=0) = P_0 / 10$ and then allow the system to evolve until the stationary solution is achieved (approximately $50\s{ps}$). To obtain the IPR of the solution we process the data in an analogous way to that done experimentally. First we normalise the wavefunction by that solution $\psi_\s{back}$ which is obtained by simulating the system without any mesas, in other words the effective `background' of the polaritons. Then we calculate the average density $\rho_n$ of polaritons in the vicinity of each mesa (labelled by $n$). We then discard those mesas located outside the pumping area, i.e. those mesas with positions $\mod{\v{R}_n} > 2\sigma$. Finally, we calculate the IPR as per Equation (\ref{IPR}) in the main text. Schematically the process looks like

\begin{enumerate}
	\item Obtain density normalised to background: $\rho(\v{r}) = \mod{\psi(\v{r})}^2 / \mod{\psi_\s{back}(\v{r})}^2$ 
	\item Calculate average density at each mesa: $\rho_n = \evalat{\langle \rho(\v{r}) \rangle}{\mod{\v{r}-\v{R}_n} < a}$ 
	\item Discard mesas outside the pumping area: $S = \{ n \in \mathbb{N}, \mod{\v{R}_n} \leq 2\sigma \}$ 
	\item Calculate IPR for the array of mesas: $\mathcal{I} = N \left( \sum_{n \in S} \rho_n^2 \right)  / \left( \sum_{n \in S} \rho_n \right)^2$
\end{enumerate}

\noindent Here, $N$ is the number of mesas used in the calculation (length of set $S$). The simulation parameters are $m = 5\times10^{-5} m_e$, $\hbar R =  0.4 \s{meV}.\mu\s{m}^2$, $\hbar \gamma  =  0.5 \s{meV}$, $\hbar g  = 2.4\times 10^{-3} \s{meV}.\mu\s{m}^2$, $V_0 = 9 \s{meV}$, $\hbar \gamma_R  = 2 \s{meV}$, and $P_0 = 2 \gamma_R \gamma / R$. Here, $V_0$ is the maxima of the trapping potential, which we model as a radially symmetric sigmoid function for each mesa.

\bibliography{sample}

\begin{thebibliography}{10}
\urlstyle{rm}
\expandafter\ifx\csname url\endcsname\relax
  \def\url#1{\texttt{#1}}\fi
\expandafter\ifx\csname urlprefix\endcsname\relax\def\urlprefix{URL }\fi
\expandafter\ifx\csname doiprefix\endcsname\relax\def\doiprefix{DOI: }\fi
\providecommand{\bibinfo}[2]{#2}
\providecommand{\eprint}[2][]{\url{#2}}

\bibitem{Anderson1958}
\bibinfo{author}{Anderson, P.~W.}
\newblock \bibinfo{journal}{\bibinfo{title}{Absence of diffusion in certain
  random lattices}}.
\newblock {\emph{\JournalTitle{Phys. Rev.}}} \textbf{\bibinfo{volume}{109}},
  \bibinfo{pages}{1492--1505} (\bibinfo{year}{1958}).

\bibitem{Segev2013}
\bibinfo{author}{Segev, M.}, \bibinfo{author}{Silberberg, Y.} \&
  \bibinfo{author}{Christodoulides, D.~N.}
\newblock \bibinfo{journal}{\bibinfo{title}{Anderson localization of light}}.
\newblock {\emph{\JournalTitle{Nat. Photonics}}} \textbf{\bibinfo{volume}{7}},
  \bibinfo{pages}{179--204} (\bibinfo{year}{2013}).

\bibitem{Dalichaouch1991}
\bibinfo{author}{Dalichaouch, R.}, \bibinfo{author}{Armstrong, J.~P.},
  \bibinfo{author}{Schultz, S.}, \bibinfo{author}{Platzman, P.~M.} \&
  \bibinfo{author}{McCall, S.~L.}
\newblock \bibinfo{journal}{\bibinfo{title}{Microwave localization by
  two-dimensional random scattering}}.
\newblock {\emph{\JournalTitle{Nature}}} \textbf{\bibinfo{volume}{354}},
  \bibinfo{pages}{53} (\bibinfo{year}{1991}).

\bibitem{Chabanov2000}
\bibinfo{author}{Chabanov, A.~A.}, \bibinfo{author}{Stoytchev, M.} \&
  \bibinfo{author}{Genack, A.~Z.}
\newblock \bibinfo{journal}{\bibinfo{title}{Statistical signatures of photon
  localization}}.
\newblock {\emph{\JournalTitle{Nature}}} \textbf{\bibinfo{volume}{404}},
  \bibinfo{pages}{850} (\bibinfo{year}{2000}).

\bibitem{Weaver1990}
\bibinfo{author}{Weaver, R.}
\newblock \bibinfo{journal}{\bibinfo{title}{Anderson localization of
  ultrasound}}.
\newblock {\emph{\JournalTitle{Wave Motion}}} \textbf{\bibinfo{volume}{12}},
  \bibinfo{pages}{129 -- 142} (\bibinfo{year}{1990}).

\bibitem{Billy2008}
\bibinfo{author}{Billy, J.} \emph{et~al.}
\newblock \bibinfo{journal}{\bibinfo{title}{Direct observation of anderson
  localization of matter waves in a controlled disorder}}.
\newblock {\emph{\JournalTitle{Nature}}} \textbf{\bibinfo{volume}{453}},
  \bibinfo{pages}{891} (\bibinfo{year}{2008}).

\bibitem{Basko2006}
\bibinfo{author}{Basko, D.}, \bibinfo{author}{Aleiner, I.} \&
  \bibinfo{author}{Altshuler, B.}
\newblock \bibinfo{journal}{\bibinfo{title}{Metal?insulator transition in a
  weakly interacting many-electron system with localized single-particle
  states}}.
\newblock {\emph{\JournalTitle{Ann. Phys.}}} \textbf{\bibinfo{volume}{321}},
  \bibinfo{pages}{1126 -- 1205} (\bibinfo{year}{2006}).

\bibitem{Gornyi2005}
\bibinfo{author}{Gornyi, I.~V.}, \bibinfo{author}{Mirlin, A.~D.} \&
  \bibinfo{author}{Polyakov, D.~G.}
\newblock \bibinfo{journal}{\bibinfo{title}{Interacting electrons in disordered
  wires: Anderson localization and low-$t$ transport}}.
\newblock {\emph{\JournalTitle{Phys. Rev. Lett.}}}
  \textbf{\bibinfo{volume}{95}}, \bibinfo{pages}{206603}
  (\bibinfo{year}{2005}).

\bibitem{Abanin2017}
\bibinfo{author}{Abanin, D.~A.} \& \bibinfo{author}{Papi{\'c}, Z.}
\newblock \bibinfo{journal}{\bibinfo{title}{Recent progress in many body
  localization}}.
\newblock {\emph{\JournalTitle{Ann. Phys.}}} \textbf{\bibinfo{volume}{529}},
  \bibinfo{pages}{1700169} (\bibinfo{year}{2017}).

\bibitem{Schreiber2015}
\bibinfo{author}{Schreiber, M.} \emph{et~al.}
\newblock \bibinfo{journal}{\bibinfo{title}{Observation of many-body
  localization of interacting fermions in a quasirandom optical lattice}}.
\newblock {\emph{\JournalTitle{Science}}} \textbf{\bibinfo{volume}{349}},
  \bibinfo{pages}{842--845} (\bibinfo{year}{2015}).

\bibitem{Choi2016}
\bibinfo{author}{Choi, J.-y.} \emph{et~al.}
\newblock \bibinfo{journal}{\bibinfo{title}{Exploring the many-body
  localization transition in two dimensions}}.
\newblock {\emph{\JournalTitle{Science}}} \textbf{\bibinfo{volume}{352}},
  \bibinfo{pages}{1547--1552} (\bibinfo{year}{2016}).

\bibitem{Kavokin2017}
\bibinfo{author}{Kavokin, A.~V.}, \bibinfo{author}{Baumberg, J.~J.},
  \bibinfo{author}{Malpuech, G.} \& \bibinfo{author}{Laussy, F.~P.}
\newblock \emph{\bibinfo{title}{{Microcavities}}} (\bibinfo{publisher}{Oxford
  University Press}, \bibinfo{address}{Oxford}, \bibinfo{year}{2017}).

\bibitem{Rodriguez2018}
\bibinfo{author}{Rahimzadeh Kalaleh~Rodriguez, S.} \emph{et~al.}
\newblock \bibinfo{journal}{\bibinfo{title}{Nonlinear polariton localization in
  strongly coupled driven-dissipative microcavities}}.
\newblock {\emph{\JournalTitle{ACS Photonics}}} \textbf{\bibinfo{volume}{5}},
  \bibinfo{pages}{95--99}, \doiprefix\url{10.1021/acsphotonics.7b00721}
  (\bibinfo{year}{2018}).

\bibitem{Abbarchi2013}
\bibinfo{author}{Abbarchi, M.} \emph{et~al.}
\newblock \bibinfo{journal}{\bibinfo{title}{Macroscopic quantum self-trapping
  and josephson oscillations of exciton polaritons}}.
\newblock {\emph{\JournalTitle{Nature Physics}}} \textbf{\bibinfo{volume}{9}},
  \bibinfo{pages}{275} (\bibinfo{year}{2013}).

\bibitem{Roumpos2010}
\bibinfo{author}{Roumpos, G.}, \bibinfo{author}{Nitsche, W.~H.},
  \bibinfo{author}{H\"ofling, S.}, \bibinfo{author}{Forchel, A.} \&
  \bibinfo{author}{Yamamoto, Y.}
\newblock \bibinfo{journal}{\bibinfo{title}{Gain-induced trapping of
  microcavity excitondensates}}.
\newblock {\emph{\JournalTitle{Phys. Rev. Lett.}}}
  \textbf{\bibinfo{volume}{104}}, \bibinfo{pages}{126403}
  (\bibinfo{year}{2010}).

\bibitem{Baboux2016}
\bibinfo{author}{Baboux, F.} \emph{et~al.}
\newblock \bibinfo{journal}{\bibinfo{title}{Bosonic condensation and
  disorder-induced localization in a flat band}}.
\newblock {\emph{\JournalTitle{Phys. Rev. Lett.}}}
  \textbf{\bibinfo{volume}{116}}, \bibinfo{pages}{066402},
  \doiprefix\url{10.1103/PhysRevLett.116.066402} (\bibinfo{year}{2016}).

\bibitem{Daif2006}
\bibinfo{author}{Da{\"i}f, O.~E.} \emph{et~al.}
\newblock \bibinfo{journal}{\bibinfo{title}{Polariton quantum boxes in
  semiconductor microcavities}}.
\newblock {\emph{\JournalTitle{Appl. Phys. Lett.}}}
  \textbf{\bibinfo{volume}{88}}, \bibinfo{pages}{061105}
  (\bibinfo{year}{2006}).

\bibitem{Jacqmin2014}
\bibinfo{author}{Jacqmin, T.} \emph{et~al.}
\newblock \bibinfo{journal}{\bibinfo{title}{Direct observation of dirac cones
  and a flatband in a honeycomb lattice for polaritons}}.
\newblock {\emph{\JournalTitle{Phys. Rev. Lett.}}}
  \textbf{\bibinfo{volume}{112}}, \bibinfo{pages}{116402}
  (\bibinfo{year}{2014}).

\bibitem{Milicevic2015}
\bibinfo{author}{Mili{\'{c}}evi{\'{c}}, M.} \emph{et~al.}
\newblock \bibinfo{journal}{\bibinfo{title}{Edge states in polariton honeycomb
  lattices}}.
\newblock {\emph{\JournalTitle{2D Materials}}} \textbf{\bibinfo{volume}{2}},
  \bibinfo{pages}{034012} (\bibinfo{year}{2015}).

\bibitem{Adiyatullin2017}
\bibinfo{author}{Adiyatullin, A.~F.} \emph{et~al.}
\newblock \bibinfo{journal}{\bibinfo{title}{Periodic squeezing in a polariton
  josephson junction}}.
\newblock {\emph{\JournalTitle{Nature Communications}}}
  \textbf{\bibinfo{volume}{8}}, \bibinfo{pages}{1329} (\bibinfo{year}{2017}).

\bibitem{Schwartz2007}
\bibinfo{author}{Schwartz, T.}, \bibinfo{author}{Bartal, G.},
  \bibinfo{author}{Fishman, S.} \& \bibinfo{author}{Segev, M.}
\newblock \bibinfo{journal}{\bibinfo{title}{Transport and anderson localization
  in disordered two-dimensional photonic lattices}}.
\newblock {\emph{\JournalTitle{Nature}}} \textbf{\bibinfo{volume}{446}},
  \bibinfo{pages}{52 -- 55} (\bibinfo{year}{2007}).

\bibitem{Wouters2007}
\bibinfo{author}{Wouters, M.} \& \bibinfo{author}{Carusotto, I.}
\newblock \bibinfo{journal}{\bibinfo{title}{Excitations in a nonequilibrium
  bose-einstein condensate of exciton polaritons}}.
\newblock {\emph{\JournalTitle{Phys. Rev. Lett.}}}
  \textbf{\bibinfo{volume}{99}}, \bibinfo{pages}{14} (\bibinfo{year}{2007}).

\bibitem{Takemura2014}
\bibinfo{author}{Takemura, N.}, \bibinfo{author}{Trebaol, S.},
  \bibinfo{author}{Wouters, M.}, \bibinfo{author}{Portella-Oberli, M.~T.} \&
  \bibinfo{author}{Deveaud, B.}
\newblock \bibinfo{journal}{\bibinfo{title}{Polaritonic feshbach resonance}}.
\newblock {\emph{\JournalTitle{Nature Physics}}} \textbf{\bibinfo{volume}{10}},
  \bibinfo{pages}{500} (\bibinfo{year}{2014}).

\end{thebibliography}

\section*{Acknowledgements}

T.S., A.B. and M.S. were supported by the Foundation for Polish Science ``First Team'' project No.\ POIR.04.04.00-00-220E/16-00 (originally: FIRST\ TEAM/2016-2/17).  M.A, M.N-T., A.A., F.Z., D.O. and M.P-O. were supported by the Swiss National Science Foundation, Project No. 153620. They thank Beno{\^i}t Deveaud for support and fruitful discussions. Numerical computations were performed with a Zeus cluster in the ACK ``Cyfronet'' AGH computer center.

\section*{Author contributions}

T.S. developed the theory, performed the simulations, processed the data, and together with M.S. wrote the manuscript. M.A. devised the project and designed the sample. A.B. and T.S. wrote the numerical code. M.N-T., A.A. and M.A. performed the experiments. A.A. developed the mesa structures. F.J. fabricated the sample. D.O. and M.P-O. supervised the experimental team. M.S. supervised the theoretical team. All authors contributed to discussions and revised the manuscript.

\section*{Competing interests}

The authors declare no competing interests.

\section*{Data availability}

All raw data and source code is available from the corresponding author upon reasonable request.

\end{document}